\documentclass[a4paper]{jpconf}
\usepackage{graphicx}
\usepackage{color}
\begin{document}
	
	\newcommand {\nc} {\newcommand}
	\nc {\beq} {\begin{eqnarray}}
	\nc {\eeq} {\nonumber \end{eqnarray}}
	\nc {\eeqn}[1] {\label {#1} \end{eqnarray}}
\nc {\eol} {\nonumber \\}
\nc {\eoln}[1] {\label {#1} \\}
\nc {\ve} [1] {\mbox{\boldmath $#1$}}
\nc {\ves} [1] {\mbox{\boldmath ${\scriptstyle #1}$}}
\nc {\mrm} [1] {\mathrm{#1}}
\nc {\half} {\mbox{$\frac{1}{2}$}}
\nc {\thal} {\mbox{$\frac{3}{2}$}}
\nc {\fial} {\mbox{$\frac{5}{2}$}}
\nc {\la} {\mbox{$\langle$}}
\nc {\ra} {\mbox{$\rangle$}}
\nc {\eq} [1] {(\ref{#1})}
\nc {\Eq} [1] {Eq.~(\ref{#1})}
\nc {\Ref} [1] {Ref.~\cite{#1}}
\nc {\Refc} [2] {Refs.~\cite[#1]{#2}}
\nc {\Sec} [1] {Sec.~\ref{#1}}
\nc {\chap} [1] {Chapter~\ref{#1}}
\nc {\anx} [1] {Appendix~\ref{#1}}
\nc {\tbl} [1] {Table~\ref{#1}}
\nc {\Fig} [1] {Fig.~\ref{#1}}
\nc {\ex} [1] {$^{#1}$}
\nc {\Sch} {Schr\"odinger }
\nc {\flim} [2] {\mathop{\longrightarrow}\limits_{{#1}\rightarrow{#2}}}
\nc {\textdegr}{$^{\circ}$}
\nc {\inred} [1]{\textcolor{red}{#1}}
\nc {\inblue} [1]{\textcolor{blue}{#1}}
\nc {\IR} [1]{\textcolor{red}{#1}}
\nc {\IB} [1]{\textcolor{blue}{#1}}
\nc{\pderiv}[2]{\cfrac{\partial #1}{\partial #2}}
\nc{\deriv}[2]{\cfrac{d#1}{d#2}}

\title{Sensitivity of one-neutron knockout of halo nuclei to their nuclear structure}

\author{Chlo\"e Hebborn}
\address{Physique Nucl\' eaire et Physique Quantique (CP 229),	Universit\' e libre de Bruxelles (ULB), B-1050 Brussels}
\ead{chloe.hebborn@ulb.ac.be}
\author{Pierre Capel}
\address{Institut f\"ur Kernphysik, Johannes Gutenberg-Universit\"at Mainz, D-55099 Mainz}
\address{Physique Nucl\' eaire et Physique Quantique (CP 229),	Universit\' e libre de Bruxelles (ULB), B-1050 Brussels}
\ead{pcapel@uni-mainz.de}

\begin{abstract}
Halo nuclei  are  located far from stability and exhibit a very peculiar structure. Due to their very short lifetime, they are often studied through reactions. Breakup reactions are of particular interest since their cross sections are large for these loosely-bound nuclei. Inclusive measurements of breakup--also called knockout reactions--have even higher statistics. In this proceeding, we study which nuclear-structure  information can be inferred from the parallel-momentum distribution of the core of one-neutron halo nuclei after the knockout of its halo neutron. In particular, we analyse the influence of the ground-state wavefunction, the presence of excited  states within the halo-nucleus spectrum and  resonances in the core-neutron continuum. Our analysis shows that such observables are sensitive to the tail of the ground-state wavefunction. The presence of excited state  decreases the breakup strength, and this flux is transferred to the inelastic-scattering channel. This indicates a conservation of the flux within each partial wave. We also show that the parallel-momentum distributions are insensitive to the existence of resonances within the continuum, they can thus be ignored in practice. This independence on the continuum argues that the parallel-momentum distributions are ideal observables to extract  very precisely the ANCs of halo nuclei.
\end{abstract}

\section{Introduction}
In the mid-eighties, the development of radioactive-ion beams  has enabled the study of nuclear structure far from stability. This has led to the discovery of unexpected features. In particular, halo nuclei have been observed~[1]. These nuclei exhibit a very large matter radius compared to stable ones. This unusual size is due to the high probability of presence of one or two nucleons far from the rest of the nucleons. They are modelled accordingly as two- or three-body objects: a compact core to which one or two nucleons are loosely bound~[2]. Archetypical examples are $^{11}\rm Be$, seen as a $^{10}\rm Be$ compact core with a one-neutron halo, and $^{11}\rm Li$, seen as a $^9\rm Li$ compact core with two neutrons in its halo. The structure of these nuclei has challenged the usual description of the nucleus as a compact object. Therefore, their study is necessary to understand the   nuclear structure close to the driplines.

Being very short-lived, halo nuclei are often probed with reactions. 
Breakup reactions describe the dissociation of the halo from the core and hence reveal the cluster structure inside the nucleus~[3].  Their cross sections are very large thanks to the fragile binding of the halo to the core. However, the detection of the neutron  reduces the statistics. Knockout reactions do not have this issue since their measurement is inclusive, i.e. only the core is detected after breakup~[4]. From a theoretical point of view, these reactions have two contributions~[5]: the diffractive breakup, where both the neutron and the core survive the collision, and the stripping part, where the neutron is absorbed by the target. They are  measured at intermediate energies between 60-100$A$~MeV and are usually analysed within the eikonal model, accurate at such energies~[5-7].  

In this work, we investigate which nuclear-structure information can be probed with one-neutron knockout reactions.  We model the  halo nucleus within the halo effective field theory (Halo-EFT)~[8,9]. This effective theory exploits the separation of scales between the  size of the compact core of the nucleus $R_{\rm core}$ and the large halo $R_{\rm halo}$ to expand the Hamiltonian at low energy. We consider the collision of $^{11}\rm Be$ with $^{12}\rm C$ at 68$A$~MeV and we focus on the sensitivity to the structure of $^{11}\rm Be$ of  the parallel-momentum distribution of $^{10}\rm Be$ after the one-neutron removal. In particular, we study its sensitivity to the $^{11}\rm Be$ ground-state wavefunction, the presence of an excited state in the $^{11}\rm Be$ spectrum and of resonances in the $^{10}\rm Be$-$n$ continuum.  The details of this analysis can be found in Ref.~[10].

Sec.~\ref{Sec2} describes briefly the reaction model and the Halo-EFT approach. The main results about the sensitivity analysis of the parallel-momentum distribution observable are in Sec.~\ref{Sec3}. In the last section, we present our conclusions. 

\vspace{-0.2cm}
\section{Theoretical framework}\label{Sec2}
We consider a projectile $P$ impinging on a target $T$ with a velocity $\ve{v}=v \ve{1}_z$. Since one-neutron halo nuclei have strongly clusterized structures, we model them as two-body objects, composed of a core $c$ and a halo neutron $n$. The structure of this projectile is  then described by the internal single-particle Hamiltonian~[11]
\begin{equation}
h_{cn}=V_{cn}(r) +  \frac{p^2 }{2 \mu_{cn}} 
\end{equation} 
where $\ve{p}$ and $r$ are, respectively, the $c$-$n$ relative momentum and distance, $\mu_{cn}$ is the $c$-$n$ reduced mass and $V_{cn}$ is an effective $c$-$n$ potential. As explained in the Introduction, halo nuclei exhibit a clear separation of scales between the size of the core $R_{\rm core}$ and the size of the halo $R_{\rm halo}$. The Halo-EFT approach~[8,9] uses this separation of scale to expand the low-energy behaviour of the $^{10}\rm Be$-$n$ system with $R_{\rm core}/R_{\rm halo} \sim 1/3$ as expansion parameter.

Following previous works~[12,13], we use the Halo-EFT at next-to-leading order (NLO) to construct the $c$-$n$ potential per partial-wave $lJ$. We parametrize it as the sum of Gaussians and their derivative
\begin{equation}
V_{cn}^{lJ} (r)=V_{0}^{lJ} e^{-\frac{r^2}{2r_0^2}} + V_{2}^{lJ} r^2e^{-\frac{r^2}{2r_0^2}},
\end{equation}
where $r_0$ represents the range of the short-range physics neglected in this model. In this work, we take it as $r_0=1.2$~fm. At NLO, we constrain the parameters $V_0^{lJ}$ and $V_2^{lJ}$ in the $s$ and $p$ waves. The $^{11}\rm Be$ halo nucleus has two bound states, $1/2^+$  and $1/2^-$, respectively  at energy  $-0.504$~MeV  and $-0.184$~MeV below the neutron separation threshold. We describe these states as single-particle states $1s1/2$ and $0p1/2$, respectively, with unit spectroscopic factors.  We fit the two parameters $V_0^{lJ}$ and $V_2^{lJ}$ in the $s1/2$ and the $p1/2$~[10]  to reproduce the experimental binding energies.  We also constrain these parameters by adjusting the single-particle asymptotic normalization constants $b_{nlJ}$ (SPANC) to the asymptotic normalization constants $\mathcal{C}_{nlJ}$ (ANC) predicted by \textit{ab initio} calculations of Calci \textit{et al.}~[14]: ${b}_{1s1/2}=\mathcal{C}_{1s1/2}=0.786$~fm$^{-1/2}$ and ${b}_{0p1/2}=\mathcal{C}_{0p1/2}=0.129$~fm$^{-1/2}$. According to the \textit{ab initio} predictions, the $p3/2$ phase shift is close to zero at low energy, therefore we do not put any interaction in this partial wave.

To describe the whole system, composed of the projectile and the target, we assume the target to be structureless and we model the $c$-$T$  and the $n$-$T$ interactions by optical potentials~[10,11]. The $P$-$T$ relative motion is given by $\Psi$, solution of the three-body \Sch equation
\begin{eqnarray}
\left[\frac{P^2}{2\mu}+h_{cn}+V_{cT}(R_{cT})+V_{nT}(R_{nT})\right]\Psi(\ve{R},\ve{r})=E_{\rm tot}\ \Psi(\ve{R},\ve{r}), \label{eq6}
\end{eqnarray}
where $\ve{P}$ and $\ve{R}$ are respectively the $P$-$T$ relative momentum and coordinate, $\mu$  is the $P$-$T$ reduced mass, $R_{(c,n)T}$  are the $c$-$T$ and $n$-$T$ distances and $E_{\rm tot}$ is the total energy of the system. We use as initial condition that the projectile is in its ground state $1s1/2$. To solve this three-body system, we use the Coulomb-Corrected eikonal approximation~[15,16]. The two contributions of the knockout cross sections, the diffractive breakup and the stripping, are computed within this model~[16,17]. In the next section, we study the sensitivity of the parallel-momentum distribution of the core after the collision to the  $1s1/2$ ground-state wavefunction,  the presence of the $0p1/2$ excited state within the projectile spectrum and the existence of a $d5/2$ single-particle resonance in the  $^{10}\rm Be$-$n$ continuum.

\vspace{-0.2cm}
\section{Results}\label{Sec3}
\subsection{Sensitivity to the ground state}
\begin{figure}[h!]
	\vspace{-0.2cm}
	\centering
	{\includegraphics[width=0.42\linewidth]{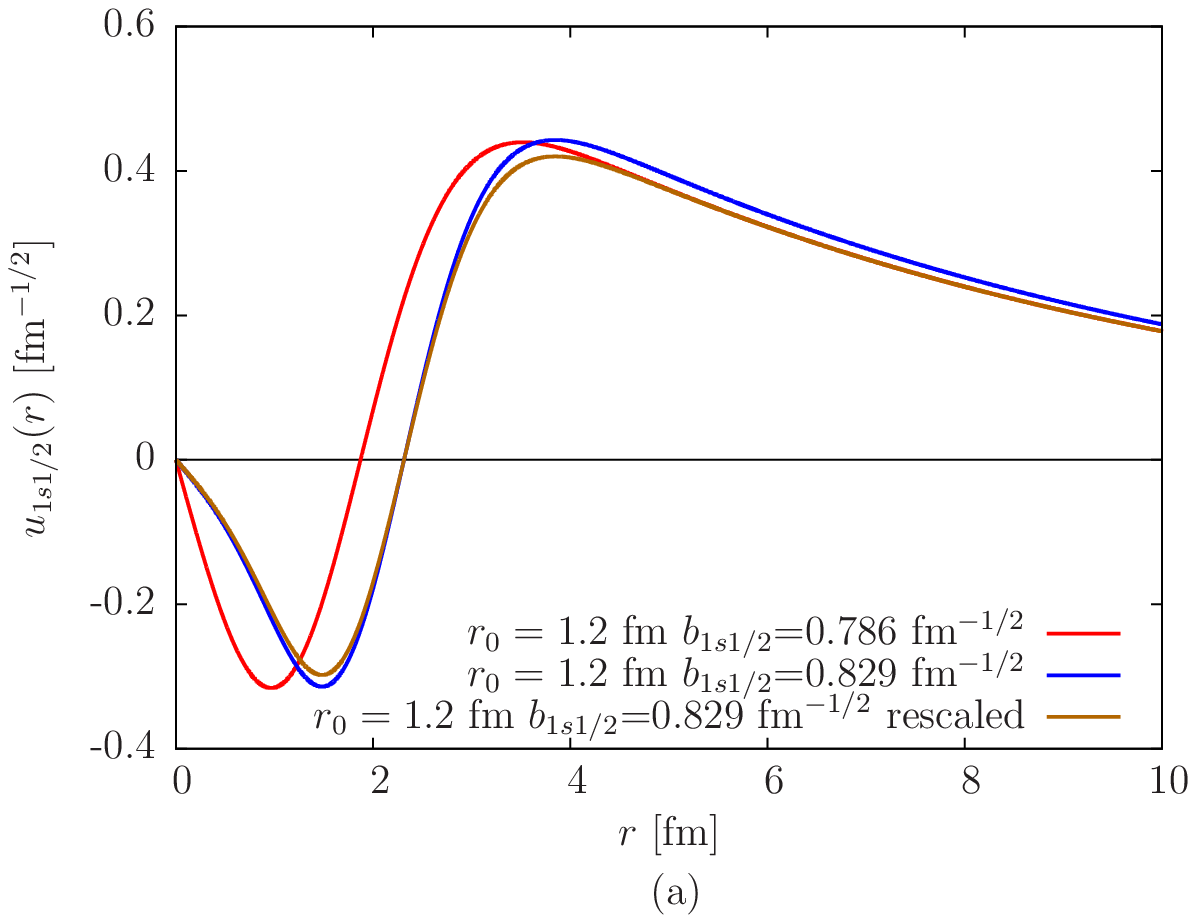}}\hspace{0.2cm}
	{\includegraphics[width=0.41\linewidth]{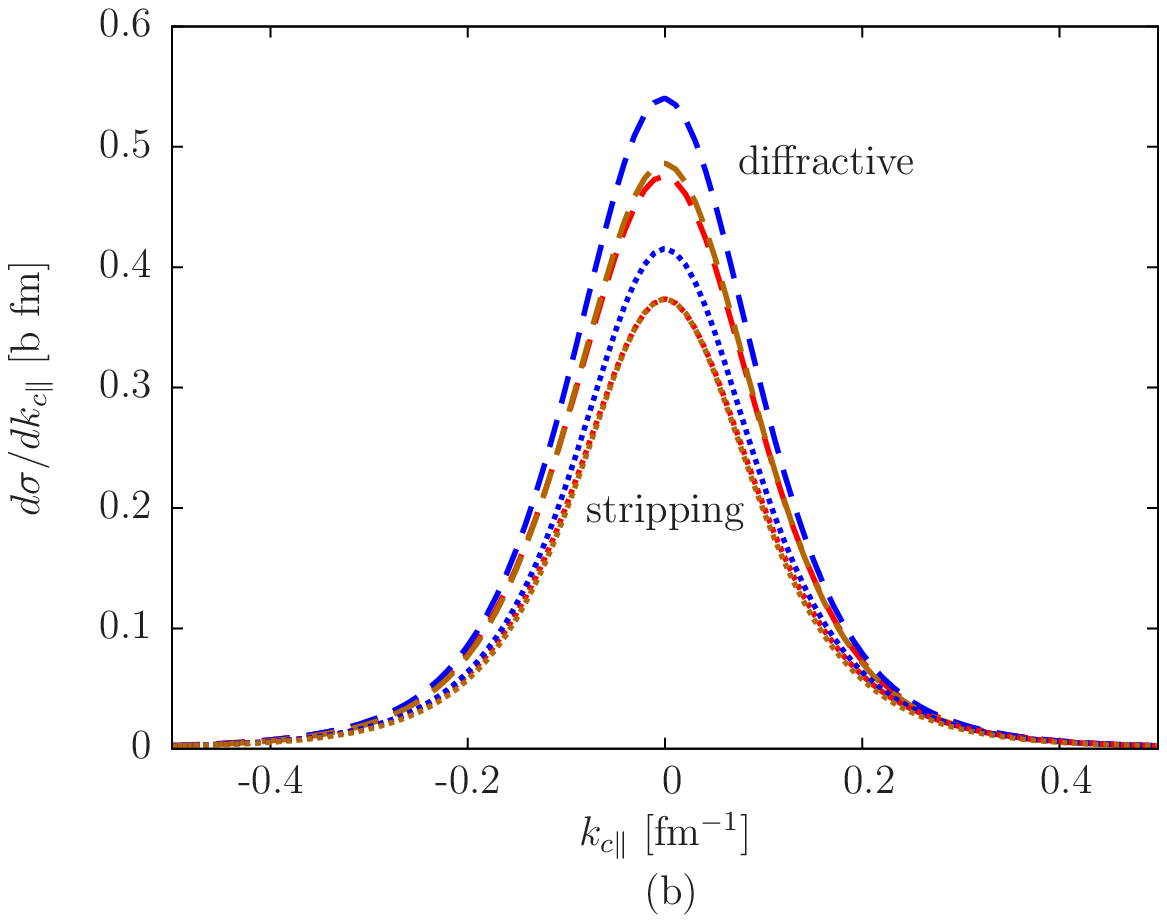}}
	
		\vspace{-0.3cm}
	\caption{(a) $^{11}\rm Be$ ground-state radial wavefunctions and (b) parallel-momentum distributions of  $^{10}\rm{Be}$ after the one-neutron knockout of $^{11}\rm Be$ with $^{12}\mathrm{C}$ at 68$A$~MeV.  The dashed and the dotted curves correspond respectively to the diffractive and the stripping contributions.}\label{FigPeriph}
\end{figure}

In Fig.~\ref{FigPeriph}(a), we plot the $^{11}\rm Be$ ground-state wavefunction obtained with the Halo-EFT potential described in Sec.~\ref{Sec2} (solid red line). 
To study the sensitivity of the knockout observable to the ground-state wavefunction, we use another Halo-EFT potential. Since the \textit{ab initio} calculations predict a spectroscopic factor of $0.9$, we fit this new potential to a larger SPANC ${b}_{1s1/2}=0.829$~fm$^{-1/2}$[$=0.786/\sqrt{0.9}$~fm$^{-1/2}$].  This leads to a new wavefunction (blue line) exhibiting a larger asymptotics and very different short-range behaviour. We then rescale this wavefunction to the  \textit{ab initio} spectroscopic factor to obtain a wavefunction (brown line) which differs  strongly below 4~fm from the one normed to unity reproducing the \textit{ab initio} ANC~[14].

\vspace{-0.2cm}
Fig.~\ref{FigPeriph}(b) displays the corresponding parallel-momentum distribution of $^{10}\rm Be$ after the one-neutron knockout of  $^{11}\rm Be$ with $^{12}\rm C$ at 68$A$~MeV. The dashed and dotted lines are respectively the diffractive and stripping contributions. One can see that the peak of the parallel-momentum distribution increases by 10~\% when the ANC of the ground-state wavefunction is larger. However, when we rescale this wavefunction to the spectroscopic factor predicted by the \textit{ab initio} calculation~[14], both the stripping and the diffractive contributions are nearly identical to the wavefunction of unit norm. This confirms the results obtained in Refs.~[18-20]: the diffractive breakup and the stripping contributions are only sensitive to the asymptotics of the ground-state wavefunction, not to its norm. The tiny discrepancy between the  diffractive distributions is due to high energies in the continuum, i.e. $E>30$~MeV, where the diffractive breakup starts to be sensitive to the inner part of the wavefunction.

This suggests that such observables could be used to reliably extract information about the tail of the wavefunction, such as the ANC of the ground state. To verify this, we check in the following the sensitivity to the presence of an excited state and to the description of the continuum.

\vspace{-0.3cm}
\subsection{Sensitivity to the excited state}
\begin{figure}[h!]
	\centering
	
	\hspace{6.2cm}{\includegraphics[clip,trim=10.5cm 2.7cm 0.7cm 5.5cm,width=0.6cm]{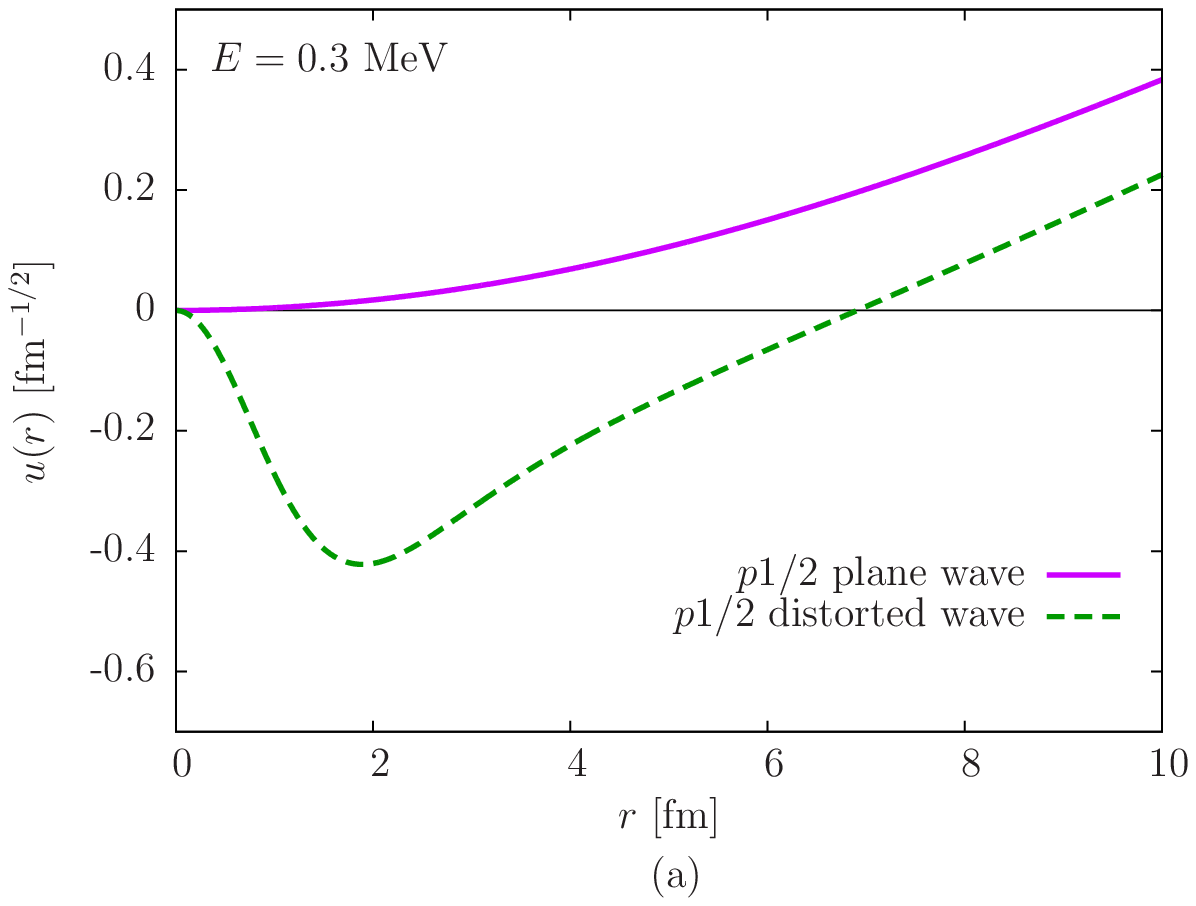}}
	
	\vspace{-0.95cm}	\includegraphics[clip,trim=0cm 0.5cm 0cm -0.1cm,width=0.45\linewidth]{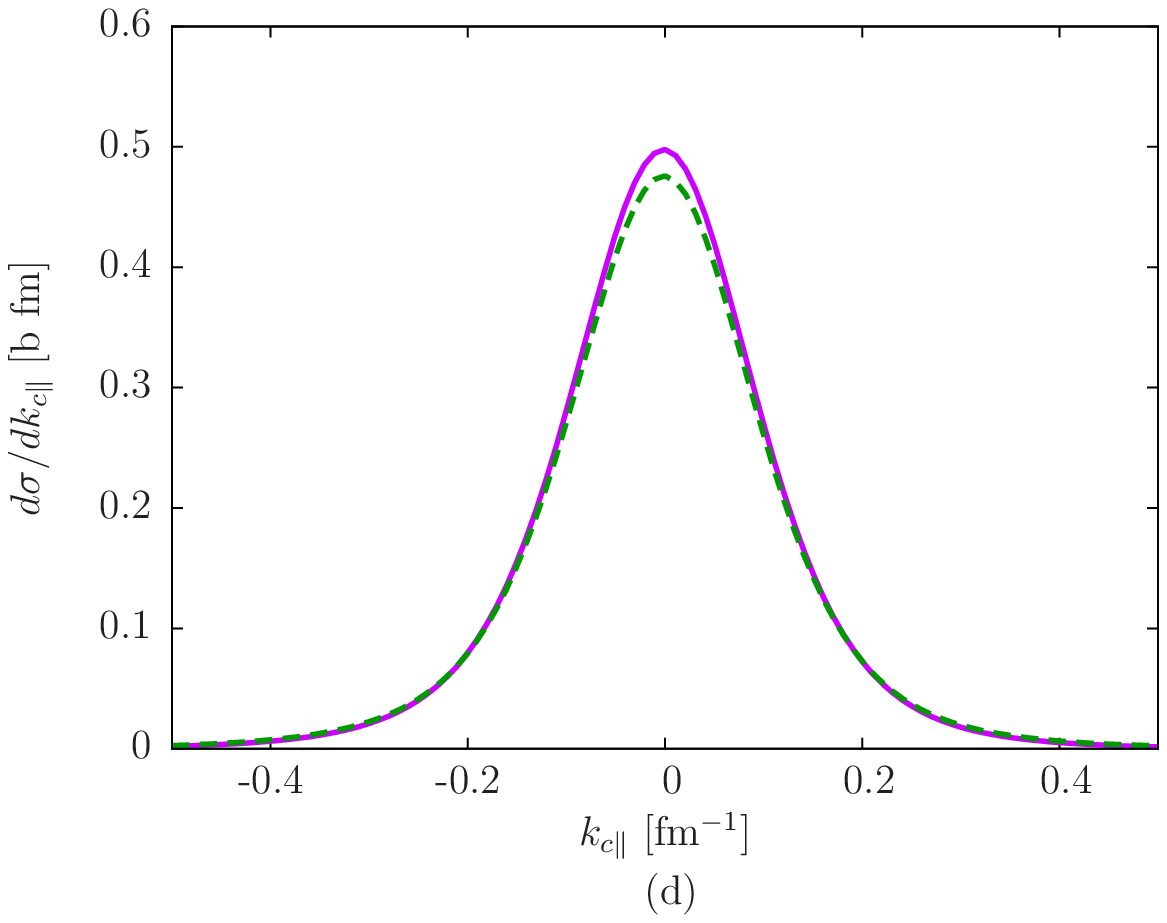}
	
	\vspace{-0.3cm}
	\caption{Parallel-momentum distribution of $^{10}\rm{Be}$ after the diffractive breakup of $^{11}\rm Be$ with $^{12}\rm C$ at 68$A$~MeV. }\label{FigExcState}
	\hspace{0.8cm}\begin{minipage}{4.5cm}\vspace{-12.25cm}\scriptsize
		
		\flushright Without the $0p1/2$ bound state
		
		\vspace{-1cm}
		\flushright With the $0p1/2$ bound state
	\end{minipage}
\end{figure}

As previously mentioned, the excited state $1/2^-$ of the $^{11}\rm Be$ is modelled through an effective interaction in the $p1/2$ wave. Such interaction also affects the $p1/2$ wave in the continuum. Since the stripping does not depend  on the excited state  or the continuum within the eikonal model~[17], we study only the parallel-momentum distribution of $^{10}\rm Be$ after the diffractive breakup of $^{11}\rm Be$ with $^{12}\rm C$ at 68$A$~MeV. To evaluate the sensitivity of this observable to the presence of the excited state, we compare in Fig.~\ref{FigExcState} the case where we do not put any interaction in the $p1/2$ partial wave (solid magenta line) to the case where we model the $0p1/2$ bound state (dashed green line). Including the $0p1/2$ excited state reduces the peak of the parallel-momentum distribution. In Ref.~[10], we show that this decrease is due to both the presence of an additional node in the $p1/2$ continuum wave, caused by the $0p1/2$ bound state, and the phase shift introduced by the interaction.

\vspace{-0.6cm}
Further analyses have shown that the presence of this bound state causes the diminution of the breakup strength within the sole $p1/2$ partial wave. In Ref.~[10], our analysis of the integrated cross sections demonstrates  that the reduction of the total breakup cross section is due to the sole presence of the $0p1/2$ bound state and not to the phase shift. This loss in the $p1/2$ breakup strength is transferred to the inelastic scattering channel, following a conservation of the flux within  each partial wave~[10].  

This study leads us to the conclusion that the excited states have a small but non-negligible impact on the parallel-momentum distribution, hence they have to be included in the analysis of the experimental data. Identical results have been obtained using the dynamical eikonal approximation (DEA)~[21,22], which does not rely on the adiabatic approximation, confirming the soundness of these results.

\vspace{-0.3cm}
\subsection{Sensitivity to the description of the continuum}
We now investigate the influence of the resonant $^{10}\rm Be$-$n$ continuum on the parallel-momentum distributions  of  $^{10}\rm Be$. As previously mentioned, the stripping cross section within the eikonal model is not sensitive to the description of the continuum~[17], therefore we  analyse only the diffractive breakup contribution. The first resonance of $^{11}\rm Be$ is a $5/2^+$ state, at $E=1.274$~MeV in the continuum with a experimental width of $\Gamma_{5/2^+}^{\rm exp}=100$~keV. To model this state, we put an interaction in the $d5/2$ to model it, going beyond the NLO of  Halo-EFT. We build various $d5/2$ potentials~[10] to produce a resonance at the energy of the $5/2^+$ state and fit its parameters to obtain different widths: $\Gamma=51$~keV, $\Gamma=98$~keV  and $\Gamma=162$~keV. 

\begin{figure}[h!]
		\vspace{-0.2cm}
	\centering
	\includegraphics[width=0.4\linewidth]{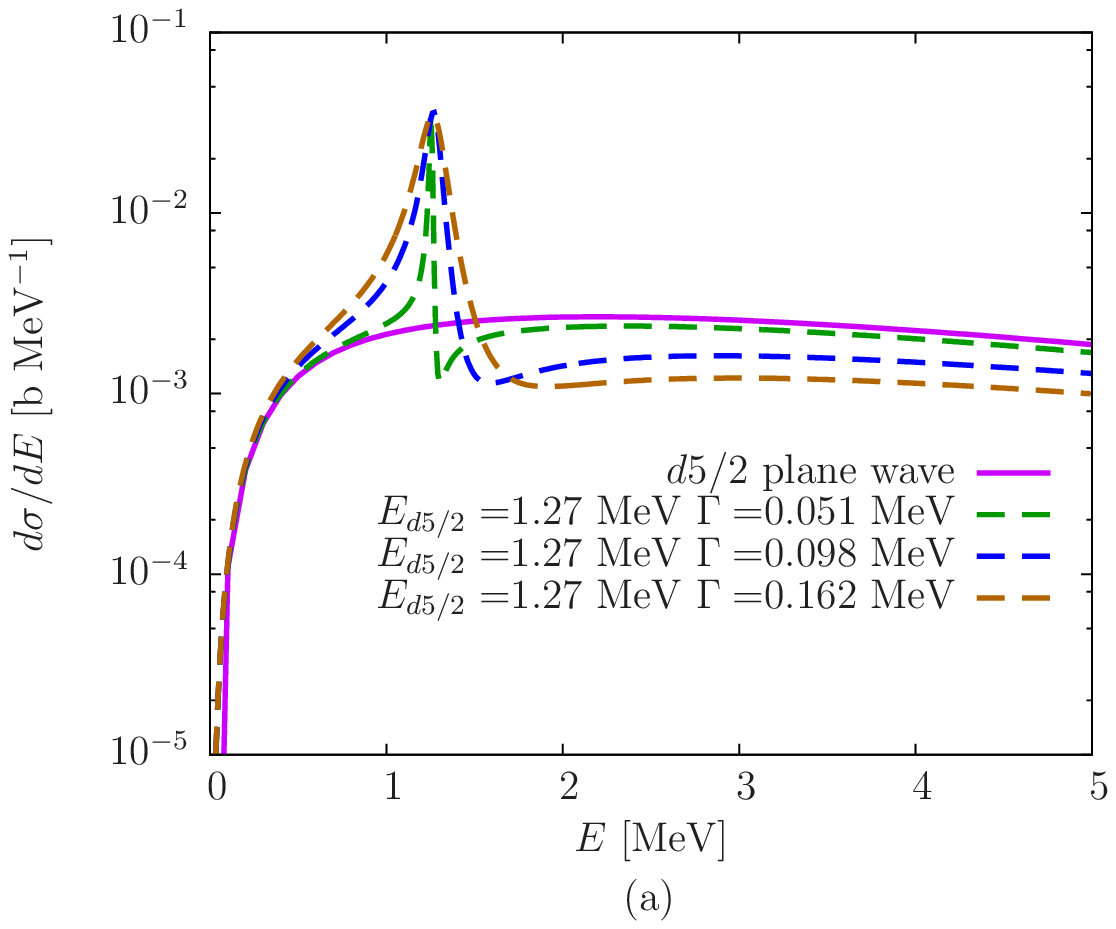}\hspace{0.2cm}	\includegraphics[width=0.42\linewidth]{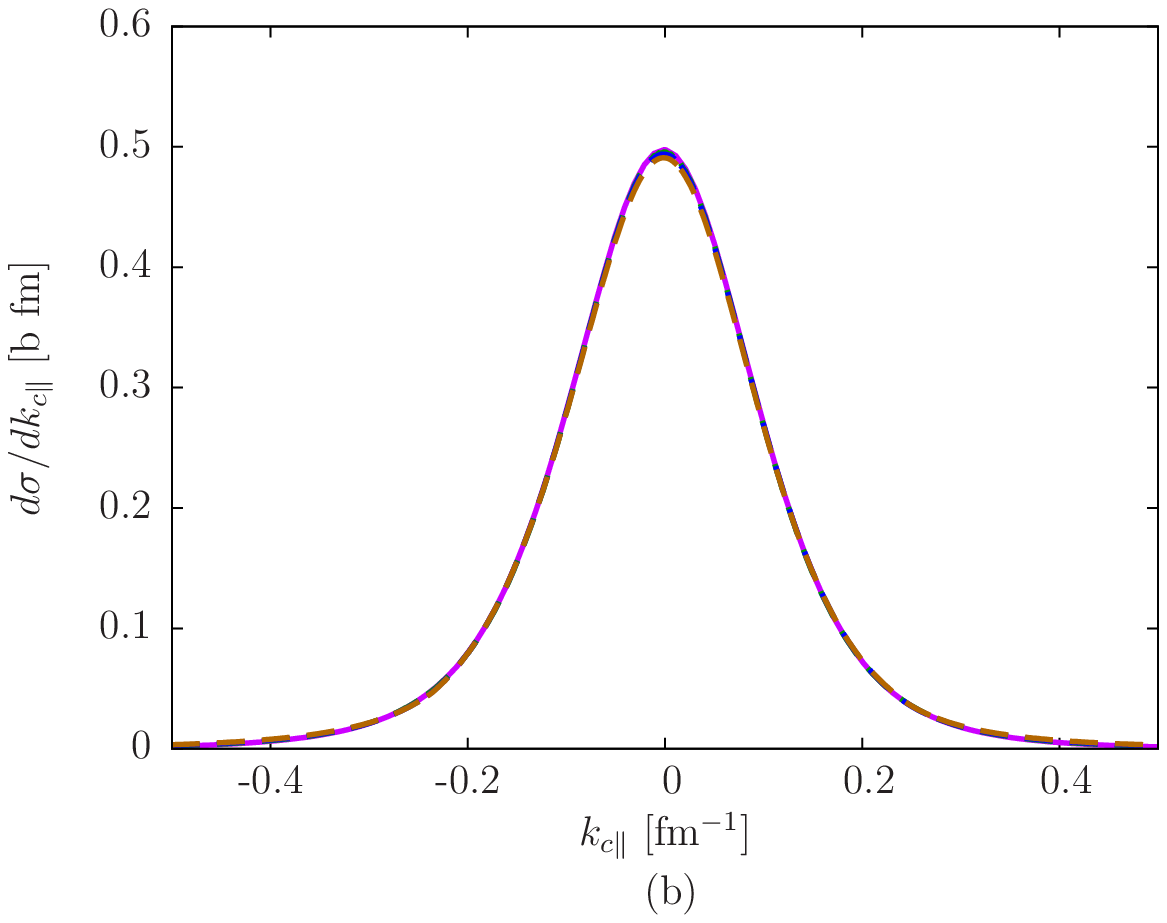}
	
		\vspace{-0.3cm}
	\caption{(a) $d5/2$ contribution to the $^{10}\rm Be$-$n$ relative energy distribution of the diffractive breakup of $^{11}\rm Be$ with $^{12}\rm C$ at 68$A$~MeV and (b) parallel-momentum of  $^{10}\rm Be$ after this breakup. }\label{FigRes}
\end{figure}

In Fig.~\ref{FigRes}(a), we plot the $d5/2$ contribution to the $^{10}\rm Be$-$n$ relative energy distribution of the diffractive breakup  of $^{11}\rm Be$ with $^{12}\rm C$ at 68$A$~MeV. We compare the case where we put no interaction in the $d5/2$ wave (solid magenta line) and where we include the $5/2^+$ states with various widths (dashed lines). As expected, in this exclusive observable, the presence of the resonance causes a large peak at the energy of the resonance, followed by a depletion area. This suppression  of the cross sections after the resonance results from destructive interferences, due to the  phase shift going above $\pi/2$. Moreover, the range of this reduction depends on the resonance width: sharper resonances lead to steeper drop but tend more rapidly to the plane-wave case.

\vspace{-0.3cm}
Contrary to the energy distributions, the  parallel-momentum distribution of  $^{10}\rm Be$  in Fig.~\ref{FigRes}(b), obtained with the different descriptions of the continuum, all superimpose. Therefore, this  observable is insensitive to the description of the resonant continuum, suggesting that describing the continuum  by plane waves is sufficient. This conclusion also holds for resonances in $p$ and $f$ waves at different energies and with various widths. As for the influence of the excited state, the same analysis with the DEA~[21,22] lead to identical conclusions. This independence on the continuum therefore pleads in favour of using the parallel-momentum distribution of the core after the one-neutron knockout to extract information about the asymptotics of the ground-state wavefunction, such as the ANC of the ground state.	

\vspace{-0.2cm}
\section{Conclusions}
Halo nuclei are often probed through reactions, such as breakup. Knockout observables have  high statistics, because the neutron is not detected after the breakup. To extract information about the halo nucleus, one needs to know precisely the sensitivity of the observable to the nuclear structure. In this work, we investigate the sensitivity of the parallel-momentum distribution of the core after the reaction to the description of the projectile: its ground-state wavefunction, the presence of a subthreshold bound state and the description of its resonant continuum. We conduct this analysis for the one-neutron knockout of $^{11}\rm Be$ with $^{12}\rm C$ at 68$A$~MeV.

Within the Halo-EFT approach, we generate various $^{11}\rm Be$ ground-state wavefunctions, having identical asymptotics but very different short-range behaviour. The comparison of the parallel-momentum distributions obtained with these wavefunctions shows that both the stripping and the diffractive contributions are sensitive only to the tail of the ground-state wavefunction. This confirms previous analyses~[18-20] and suggests that only information pertaining to the asymptotics of the ground-state wavefunction, such as the ANC, can be reliably extracted from such observables.

Our analysis also emphasizes that the presence of excited states decreases the breakup strength, and hence the magnitude of the peak of the parallel-momentum distribution. Since this loss of flux is transferred to the inelastic scattering channel, we argue for the existence of a closure relation within each partial wave~[10]. The excited states have to be included in the analysis. On the contrary, the parallel-momentum distributions are insensitive to the resonances in the continuum, therefore describing the continuum  by plane waves is sufficient. This independence on the continuum strongly reduces the uncertainty related to the description of the halo nucleus.

A direct application of this work, is the reanalysis of experimental data on $^{11}\rm Be$ and $^{15}\rm C$ one-neutron halo nuclei within the Halo-EFT framework. We plan to compare the  ANC inferred from these analyses to the ones predicted by \textit{ab initio} calculations~[14]. 

\vspace{-0.2cm}
\section*{Acknowledgments}
	C.~Hebborn acknowledges the support of the Fund for Research Training in Industry and Agriculture (FRIA), Belgium.   This project has received funding from the European Union’s Horizon 2020 research	and innovation program under grant agreement 	No 654002, the Deutsche Forschungsgemeinschaft within the Collaborative	Research Centers 1245 and 1044, and the PRISMA+ (Precision Physics, Fundamental Interactions and Structure of Matter) Cluster of Excellence. P. C. acknowledges the support of the State of Rhineland-Palatinate.

\end{document}